# Coalescence of Drops Near A Hydrophilic Boundary Leads to Long Range Directed Motion

Manoj K. Chaudhury*, Aditi Chakrabarti and Tapasya Tibrewal
Department of Chemical Engineering
Lehigh University, Bethlehem, PA 18015

**Abstract.**

A new mechanism for the passive removal of drop on a horizontal surface is described that does not require pre-fabrication of a surface energy gradient. The method relies upon the preparation of alternate hydrophilic/hydrophobic stripes on a surface. When one side of this surface is exposed to steam, with its other surface convectively cooled with cold water, steam condenses as a continuous film on the hydrophilic stripes but as droplets on the hydrophobic stripes. Coalescence leads to a random motion of the center of mass of the fused drops on the surface, which are readily removed as they reach near the boundary of the hydrophobic and hydrophilic zones thus resulting in a net diffusive flux of the coalesced drops from the hydrophobic to the hydrophilic stripes of the surface. Although an in-situ produced thermal gradient due to differential heat transfer coefficients of the hydrophilic and hydrophobic stripes could provide additional driving force for such a motion, it is, however, not a necessary condition for motion to occur. This method of creating directed motion of drops does not require a pre-existing wettability gradient and may have useful applications in thermal management devices.

**Keywords**: Hysteresis, Directed Motion, Coalescence.

*email: mkc4@lehigh.edu

# 1. Introduction.

Liquid drops move on a flat surface if it is subjected to a surface tension driven unbalanced force. A widely studied method to generate such a motion is to place a drop on a surface that has a gradient of wettability [1-7]. Due to the difference in the intrinsic contact angles on two opposite sides of the drop, a curvature gradient is imposed upon it that causes a net motion of the drop from the less to the more wettable region of the surface. Recently, it has been found that a hydrophobic surface having a morphological [8-10], a curvature gradient [11,12] and electro-wetting [13-16] also induce such types of motions. Liquid drops can also move on a surface if a thermal gradient [2,17-19] is imposed that induces a motion due to unbalanced surface tension on the liquid surface and the drop usually moves from the hotter to the colder part of the gradient. Reactive spreading [20-23] is another mechanism of drop motion that relies upon the in-situ created surface energy gradient on a surface.

The main detriment to the motion of liquid drops is related to hysteresis [3,7,24] of the type that pins the edge of the drop on any real surface. While it is not practical to produce a defect free surface, hysteresis can, however, be advantageous if the surface is subjected to a mechanical vibration [25-32]. The role of hysteresis can be viewed in two ways. Hysteresis on a heterogeneous surface can be asymmetric [27], which, like a diode, rectifies an externally imposed structured noise that gives rise to a unidirectional motion. When a random noise is imposed, the non-linear friction due to hysteresis manifests in terms of a quasi-linear friction with an emergent relaxation time depending upon the noise strength and the magnitude of hysteresis [33-36]. The effective relaxation time ($\tau_L^*$) of the surface can be expressed [36] in terms of the Langevin relaxation time ($\tau_L$) and the hysteresis ($\Delta$) as:

$$\frac{1}{\tau_L^*} = \frac{1}{\tau_L} + \frac{\Delta^2}{K} \tag{1}$$

With the above effective relaxation time, we can also define an effective diffusivity as $D \sim K\tau_L^{*2}$, which assumes the form $D \sim K^3/\Delta^4$ in the limit of low noise strength. In the presence of an externally applied force $\bar{\gamma}$, the drift velocity of the drop [36] is:

$$V_d = \bar{\gamma}\tau^* = \frac{\bar{\gamma}\tau_L}{1 + \Delta^2\tau_L/K} \tag{2}$$

In equations (1) and (2) both the hysteresis force $\Delta$ and the externally applied force $\bar{\gamma}$ are expressed on the basis of unit mass of the drop, which, thus have the units of acceleration. The noise strength is defined as: $K = \langle \gamma^2(t) \rangle \tau_c$, where $\gamma(t)$ is the value ($m/s^2$) of the noise pulse and $\tau_c$ is its duration (40 $\mu$s). $K$ is a measure of power input per unit time.

The bias can be provided either by a chemical or a thermal gradient of surface energy. The force due to the chemical gradient is $\gamma_w \left(\frac{d\cos\theta}{dx}\right) A$, where $\gamma_w$ is the surface tension of water drop, $A$ is its base area and $d\cos\theta/dx$ is the wettability gradient. When a liquid drop on a surface is under the influence of a thermal gradient, the main driving force arises from the gradient of surface tension of the liquid due to a gradient of temperature (*T*), which results in a Marangoni flow on the surface [2] of the liquid drop causing it to move towards the colder side of the gradient. The magnitude of this force is $\sim \left(\frac{d\gamma_w}{dT}\right)\left(\frac{dT}{dx}\right) A$, although this force would be randomized when subjected to vibration [19]. In both cases, however, the drop experiences a resistive force due to wetting hysteresis as its frontal side attempts to reach the local advancing angle, whereas its rear side tends to attain a local receding angle. This leads to a threshold force of magnitude: $\gamma_w b(\cos\theta_r - \cos\theta_a)$, where $b$ is its width, and $\theta_a$ and $\theta_r$ are the advancing and receding contact angles on a given location on the surface respectively. Equation (2) predicts that the drift velocity of the drop increases with the noise strength till a limiting velocity is reached that is controlled only via Langevin relaxation time [36] (see also the Appendix). In the case of drops condensing on a surface, a self-generated noise can induce a random motion of drops, the strength of which depends on the excess surface energy due to random coalescence.

Controlled transport of drops has many practical applications in biology [37], as well as in unit operations involving water and thermal managements [4]. Thermal management is of considerable importance in micro-heat exchanger and heat pipe technologies. Vapor condenses as a thin continuous film on a hydrophilic surface that reduces the heat flux. While dropwise condensation can enhance the heat flux and external force is needed to remove them. While the dropwise condensation on an inclined surface can enhance heat flux on the vapor side of a heat exchanger, it is not effective on a horizontal surface as the drops coalesce and grow upon it to cover the entire surface. These issues were discussed in a previous publication [4], where we demonstrated that the heat flux through a horizontal metal disc, one side of which is convectively cooled while the other side of which is exposed to steam, can be enhanced [4] if a wettability gradient is designed on the surface of the disc that is in contact with the steam. As the steam condenses on such a gradient surface, the droplets are removed from the center to the edge of the disc by the gradient that enhances the net heat flux much more than that on a hydrophilic surface where steam condenses as a film. In the removal of the drops from a gradient surface, coalescence [4,38,39] of the drops plays a very important role. On a homogeneous surface, the coalescence of two equal size drops causes the free edges of the drops to move towards their equilibrium position as the surface area and thus the surface energy of the coalesced drops is minimized. If two asymmetric drops coalesce, there is a net movement of the resulting center of mass on the surface, which results in a random motion that is reminiscent of a self-avoiding random walk. When such coalescence occurs on a surface possessing a mechanism to break the symmetry, a directed motion of the drops occurs on the surface. Combination of a heterogeneous wettability and hierarchical roughness can also be used to direct motion of drop on a condensing surface as was demonstrated in a recent nice study [40]. Coalescence plays an important role here as well. Our current paper draws inspiration from motion generated in out of equilibrium systems exhibiting large fluctuations [36] in that the random coalescence of drops gives rise to a self-generated noise, the strength of which depends upon the degree of sub-cooling of the substrate, which, in turn, affects the density and the rate of nucleation as well as the random velocity of the motion of the coalesced drops. The similarity of a drop moving on a gradient surface due to an external noise and the heat flux resulting from the motion of the coalesced drops on a gradient surface is illustrated in the Appendix.

So far, a chemical or a morphological gradient has been explored to effectively remove condensed drops on a surface. Here we propose a new mechanism to remove condensed drops, the principle of which is illustrated in figure 1. Consider a flat surface that has alternate stripes of hydrophilic and hydrophobic zones. Steam condenses as a film on the hydrophilic stripes but as drops on the hydrophobic stripes.

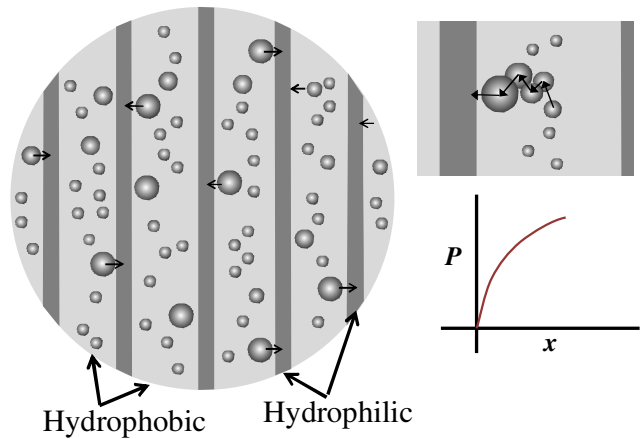

**Figure 1**. When a surface consisting of alternate stripes of hydrophobic/hydrophilic zones is exposed to steam while its other side is cooled convectively, the steam condenses as thin film on the hydrophilic stripes, but as drops on the hydrophobic stripes. The right image shows a zoomed portion of the patterned silicon wafer showing the random motion of drops due to coalescence. As the coalesced drops reach near the hydrophilic stripe (the hydrophilic stripe can be considered as a well, where probability of finding a drop is zero), it merges with the liquid film collected on this stripe. The graph is an idealized depiction of the probability density of drops as a function of the distance.

As the drops coalesce on the hydrophobic surface, the center of mass of the fused drops undergoes a random walk. When these drops occasionally come in contact with the edge of the condensed water film, they are readily pulled into the hydrophilic zone due to the difference in the Laplace pressure of the drop and the film. This scenario is remotely similar to a single component diffusion of a species near an absorbing boundary, in that a gradient of the concentration is created near the boundary of the two zones that creates a net diffusive flux of the coalesced drops from the hydrophobic to the hydrophilic sides of the surface. Once the liquid is collected on the hydrophilic channels, it can be removed by a wicking mechanism that can be designed on the steam side of a metal disc. In this paper, we intend to provide a proof of the concept based on the evidences reported below.

2. **Experimental Details**.

## 2.1. Preparation of striped surface.

In order to create a hydrophobically modulated striped pattern on the silicon wafer, a plastic template is formed by cutting out thin stripes from a regular transparency sheet (0.12mm thick, 3M) (Figure 2). After exposing a silicon wafer (0.29 mm thick, 2" diameter, Silicon Quest International) to a high temperature flame of a propane torch to remove any organic impurities, it is cooled and placed on the base of a polystyrene petri dish (VWR, 150 mm diameter, 10 mm high). The patterned template was placed above the wafer. A filter paper soaked with a few drops of decyltrichlorosilane (Gelest Inc.) was attached underneath the upper lid of the petri dish and used to cover its base containing the sample. After exposing the silicon wafer to the vapor of the silane for one minute, the upper lid was removed. The silane reacted with the exposed area of the wafer, but not with the area that was covered. This way the silicon wafer possessed alternate stripes of hydrophobic/hydrophilic regions on its surface. Although we have been performing experiments with different ratios of hydrophilic/hydrophobic surface areas, here we report the data obtained with a surface, 25% (3mm each) of which is hydrophilic and 75% (9 mm each) of which is hydrophobic.

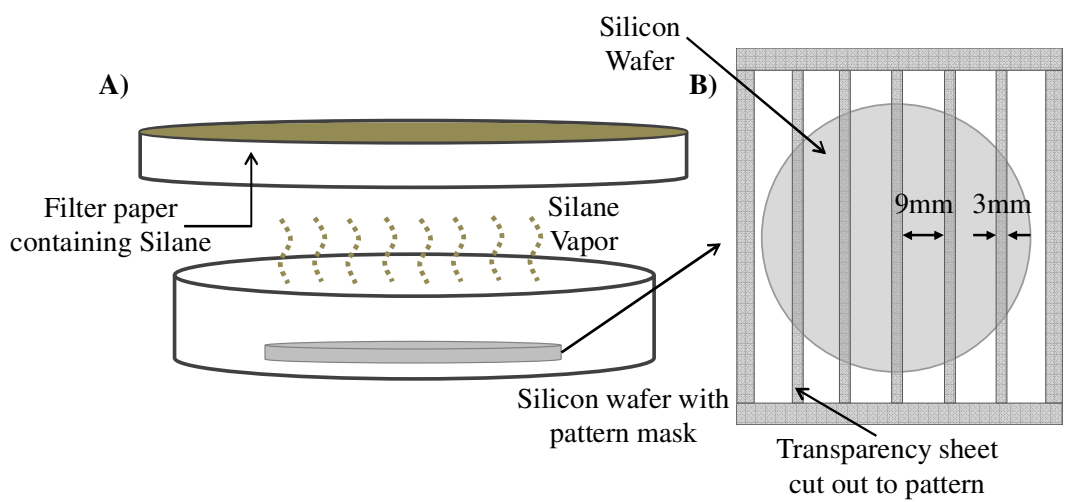

**Figure 2.** A) Schematic of the method used to prepare hydrophilic/hydrophobic stripes on a silicon wafer. After placing a thin plastic template on the silicon wafer, it was exposed to the vapor of decyltrichlorosilane in a closed petri dish. B) Plan view of the Silicon wafer with the plastic template on it. The whole assembly is placed on the base of the petri dish in which the vapor deposition of Silane is carried out. The exposed part was hydrophobic, whereas the unexposed part remained hydrophilic.

**2.2. Design of the heat exchanger incorporating patterned Silicon wafer.**

The basic heat exchanger [4,41] is a horizontal copper cylinder encased inside a Teflon jacket as we have described in the past. The silicon wafer, upon which alternate stripes of hydrophobic and hydrophilic zones are designed, is pressed in contact with the upper surface of copper with the help of two metal clips that snugly fit around the side of the Teflon jacket. The patterned silicon wafer, on copper cylinder, comes in contact with the steam, whereas the other flat end of the cylinder is convectively cooled by water. This design ensures [4,41] unidirectional flux of heat along the axis of the cylinder.

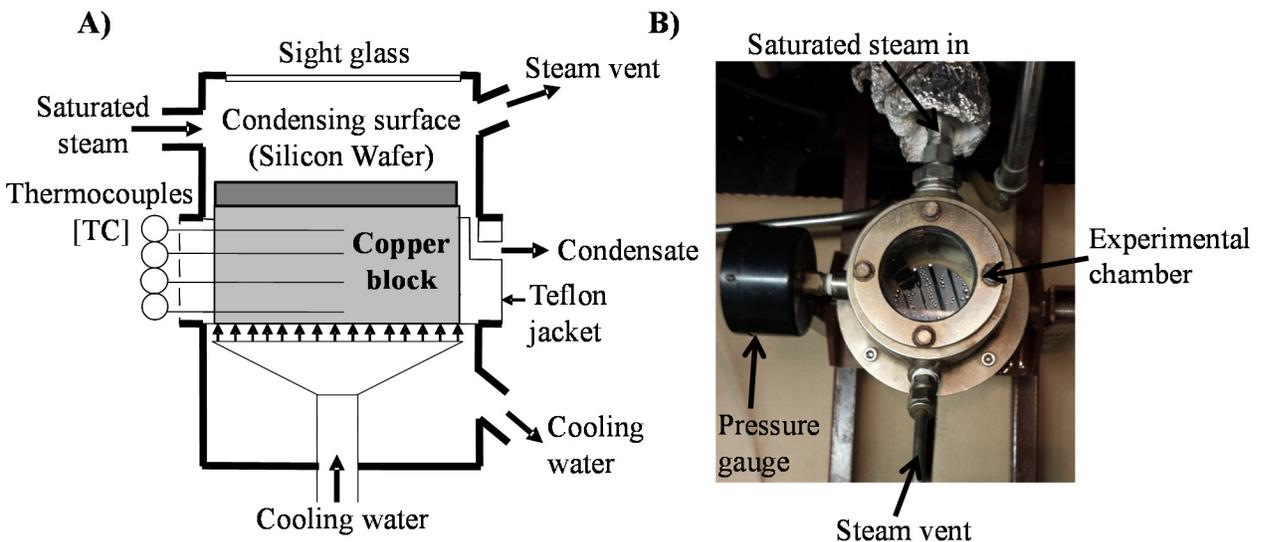

**Figure 3.** A) Schematic of a heat exchanger used to measure heat flux through a horizontal surface of a silicon wafer tightly pressed against a copper block. Steam passes over the upper surface and condenses over the wafer as the other side of the copper block is cooled by convective flow of water as shown. The four thermocouples placed at different axial positions of the copper block allow estimation of the temperature gradient, which when multiplied by the conductivity of the copper yields the heat flux. B) A plan view of the apparatus showing film-wise condensation of steam on the hydrophilic stripes, whereas dropwise condensation on the hydrophobic stripes.

The entire assembly is then placed inside a metal chamber which has a sight glass at its top to view the condensation phenomena. Fine holes are laser drilled through the side of the copper cylinder that allows insertion of four thermocouples (figure 3A), thereby enabling measurement of the thermal gradient along the axis of the cylinder during the heat transfer process.

Steam generated (3.6 kg/hr) by a steam generator (Sussman, Model: MB3L) was passed through the upper surface of the silicon wafer, whereas the lower surface of the copper block was convectively cooled with cold water as shown in figure 3A. By controlling the steam and the water flow rates, the degree of the sub-cooling of the steam was varied. A regular camera (Samsung) was used to capture the sequence of events.

## 2.3. Impingement of an atomized water jet on the patterned Si wafer to observe coalescence of drops.

The patterned Silicon wafer containing alternate hydrophobic (75%) / hydrophilic (25%) stripes was mounted vertically in ambient conditions. An atomized jet of deionized water was impinged on its surface at 0.44 ml/s with the help of a home built atomizer consisting of a thin stream of water being directed towards the surface with a continuous stream of air. The sequence of events were captured by a regular camera.

## 3. Results and Discussions.

The striped surface was characterized by depositing droplets (1 $\mu$l) of water across the hydrophilic/hydrophobic zones. Although the drops spread with non-measurable contact angles on the hydrophilic stripe along the length of the stripe, they are confined at the boundaries of the hydrophilic/hydrophobic stripes. In between two hydrophilic stripes, an average contact angle of about 93° was observed. As shown in figure 4, the contact angles of four different drops placed on the hydrophobic stripe are similar, thus suggesting that no gradient of wettability exists on the hydrophobic stripe.

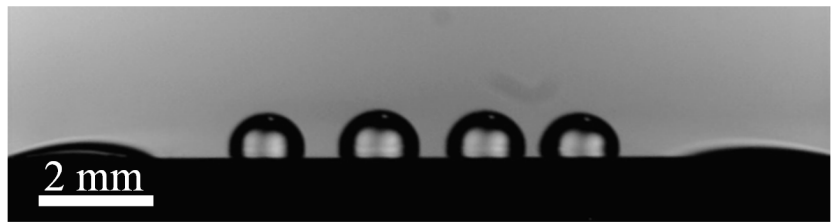

**Figure 4**. Four 1 µl size water drops placed on a hydrophobic stripe show the uniformity of contact angles (93°). Two hydrophilic channels on the two sides of this hydrophobic stripe are hydrophilic, which are wetted by water.

When steam was passed over such a striped silicon wafer with its other surface cooled by a convective flow of cold water (19 °C ), it condensed as thin film within the hydrophilic stripes, but as drops on the hydrophobic stripes., A camera shot of the plan view of the experimental assembly (Figure 3B), shows such stripes that become decorated by differential condensation on the hydrophilic/hydrophobic stripes.

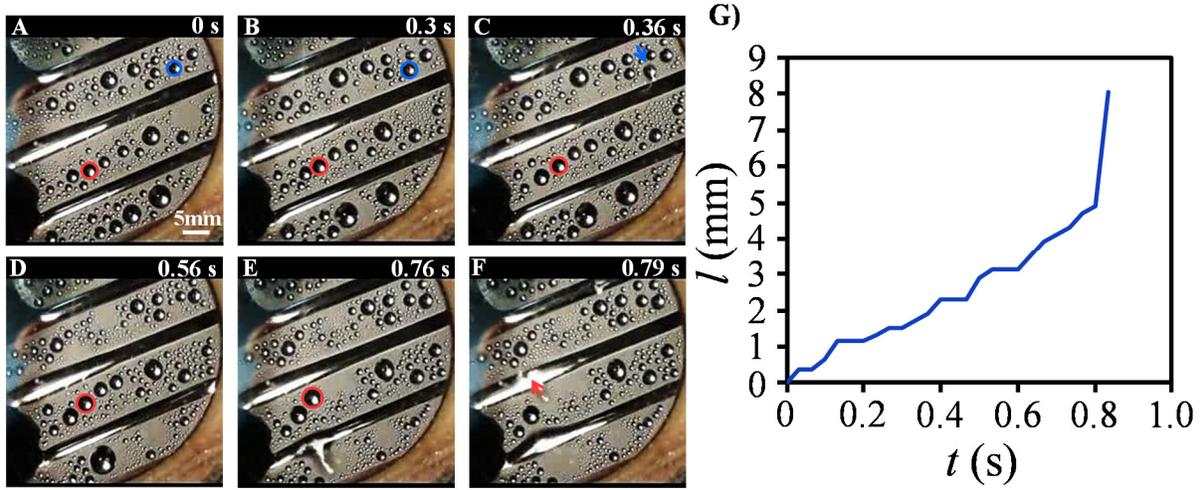

**Figure 5.** (A-F) Video snapshots of the sidewise motions of the condensed water drops from the hydrophobic to the hydrophilic stripes when the heat exchanger is in operation. Two drops are circled to illustrate their motions on the opposite directions. (G) Distance traversed by the drop circled with red in the snapshots measured at equal intervals of time, i.e. 0.033s.

When stream continues to flow through the chamber, a temperature gradient develops in the copper block along the axial direction. The control parameter that determines the heat flux is the degree of sub-cooling ($\Delta T$) that is the difference between the steam temperature and the average surface temperature of the silicon wafer, which was estimated by measuring the gradient of the temperature along the axis of the copper block and extrapolating it to the surface. The heat flux itself was estimated in two ways. One method estimates the heat flux from $k_{Cu}\, dT/dx$, where $k_{Cu}$ is the thermal conductivity of the copper. The other method uses the measurement of the condensate collected from inside the thermal chamber using a wick and using the formula $\dot{m}H_m$ where, $\dot{m}$ is the condensate flow rate and $H_m$ is the latent heat of vaporization of water per unit mass. Although both the methods yielded similar values of heat flux, we use here the values obtained directly from the temperature gradient measured along the axis of the cylinder.

Figure 5 captures a few video snapshots of the silicon wafer that shows that steam condenses as a thin film of water on the hydrophilic zone, but as drops on the hydrophobic zone. The sidewise motion of the drops from the hydrophobic to the hydrophilic zone is also evident (See Movie 1 in Supporting Information). As the drops coalesce undergoing a random walk, they occasionally come in contact with the edge of the condensed water film. When that happens, the drops are readily pulled into the hydrophilic zone due to the difference in the Laplace pressure of the drop and the film. The velocities of the coalesced drop depend on its size and, thus, vary. The trajectory of a typical drop is shown in figure 5G, where it is observed that the drop moves a short distance by coalescence with other small drops. When coalescence does not occur, the drop stays stationary till it coalesces with another drop growing in close proximity. After a series of stop/go events, the coalesced drop comes close to the edge of the hydrophobic/hydrophilic zones where, aided by rapid coalescence of other drops, a faster motion of the drop occurs till it merges with the water film on the hydrophilic stripe. The average velocity of this drop is about 1 cm/s.

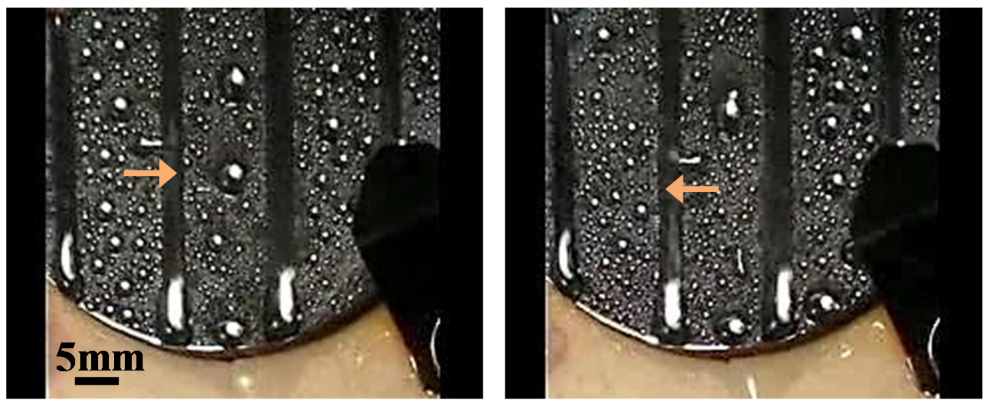

**Figure 6**. Random snapshots captured from a video showing the directed motion of water droplets (as streak of lights) from hydrophobic to hydrophilic stripes when a mist of water drops are produced on the surface at room temperature. The arrows indicate the direction of the drops when they get pulled into the hydrophilic regions.

Even though the above model of random walk with an absorbing boundary is the main mechanism (see below) for the motion of the drops observed in the above experiment, it is possible that an in-situ produced thermal gradient parallel to the surface of the condenser could play an additional role. To illustrate this point, let us consider that the temperature of the surface just underneath the water film condensed on the hydrophilic stripe is $T_1$ and that of the hydrophobic surface is $T_2$. Let us consider that the heat transfer coefficient of the water film is $h_1$ and the interfacial resistance from the steam to the hydrophobic surface (contributed by the vapor layer and the condensed droplets) is $h_2$. Let us compare $T_1$ and $T_2$, by assuming that the net heat fluxes through the hydrophilic and the hydrophobic stripes are the same. Rigorously, this is not an exact assumption, but it yields a first order estimate of whether $T_1$ is greater, less or equal to $T_2$. Thus we set $h_1(T_{steam}-T_1) \sim h_2(T_{steam}-T_2)$, which leads to $T_2 \sim (h_1/h_2)T_1+(1-h_1/h_2)T_{steam}$. As $h_1<h_2$, we expect that the temperature $T_2$ to be closer to $T_{steam}$ and thus greater than $T_1$. As a sharp boundary cannot exist, we may expect that a temperature gradient could be set up along the surface near the boundary of the two stripes. The condensed drops could thus move from the hydrophobic to the hydrophilic stripe aided by this gradient, which is the usual Thermal Marangoni effect. Although such a temperature gradient could lead to a rectified motion of the condensed drops, we found it difficult to quantify this effect as it requires direct measurement of the temperature gradient parallel to the surface close to the boundary of the two stripes. Remote sensing of the surface temperature using an infrared camera was not feasible as the steam above the surface intervened such a measurement. Interrogation with a contact measurement device is also not suitable as it modifies local heat transfer properties. The backside of the wafer is also in contact with the copper disc that is convectively cooled by water which prevented measurement from the backside of the wafer. Placement of the thermocouples at different locations of the radial direction of the copper block could not detect any variation of temperature, suggesting that the temperate gradient, even if it may be present on the surface, vanishes along the axis of the block. We, however, were able to demonstrate using a very different experiment that the existence of a lateral temperature gradient on the surface is not a necessary condition for the drops to exhibit the type of motion reported as above.

In order to demonstrate that random coalescence with an absorbing boundary is the main driving force behind the type of motion observed here, an experiment was performed in which a gentle mist of water drops was directed towards a striped surface (this time placed vertically) of the type used for the heat exchanger applications under an ambient condition. Directed motion of drops was found to occur from the hydrophobic to the hydrophilic stripes of the wafer, whereas the large drops were drained by gravity. We believe that this motion is the result of the diffusion of the coalesced drops from the central region of the gradient towards the boundary of the hydrophobic and hydrophilic zones, where they are removed by the hydrophilic stripes (See Movie 2 in Supporting Information).

The detailed solution of the problem of coalescence induced drop motion elicits a statistical mechanical description of self-avoiding random walk with absorbing boundary conditions by building upon a previous study of Marcos-Martin et al, who demonstrated the self-diffusive nature of coalesced drops [42]. However, couple points are, clear from the outset. As is the case with any diffusion problem, the flux of the coalesced droplets is proportional to a diffusivity and a gradient of the probability density of fused drops. At a low noise strength, as diffusivity ($D$) of the drop is mainly controlled by hysteresis ($\Delta$) and the strength of the noise ($K$), we expect the diffusivity ($D \sim K^3/\Delta^4$) to be a strong function of the noise strength [33-36]. As the diffusivity strongly depends on noise strength, which in turn depends on the rate of coalescence, it can be adjusted over a wide range of noise strength. Since the noise strength is a measure of power input per unit mass, using dimensional argument, it assumes the form $K = (\gamma_w k_w \Delta T / m H_v)$ for the condensing vapor, where $\gamma_w$, $k_w$ and $H_v$ are the surface tension, the thermal conductivity, and the latent heat of vaporization of water per unit volume; $m$ is the mass of an average size of the water drop on the hydrophobic surface and $\Delta T$ is the degree of subcooling, that is equal to $T_{steam}-T_2$. The physical meaning of this dimensional relation is as follows. Noise results from the excess surface energy resulting from the coalescence of two drops, which imparts energy for the coalesced drop to fluctuate; thus it is directly proportional to $\gamma_w$. Both $k_w$ and $\Delta T$ increase the heat transfer rate, that enhances the growth rate of the drop. With higher latent heat ($H_v$), only smaller rate of condensation is needed for a given heat flux controlled by $k_w$ and $\Delta T$. The unit of this dimensional group is $m^2/s^3$.

This non-linear diffusivity, in conjunction with the gradient of the probability of coalescence create a net diffusive flux in such a way that all the drops delivered to the on the hydrophobic part of the surface is transported to the hydrophobic channel. Thus the system is expected to operate with a high efficiency, which would be of significant use in various water management and heat transfer technologies. This method also has the advantage of the ease of preparation, in that any regular metal surface of a heat exchanger could simply be printed with appropriate coatings in order to generate alternate hydrophilic/hydrophobic patches for efficient removal of condensed vapors from its surface.

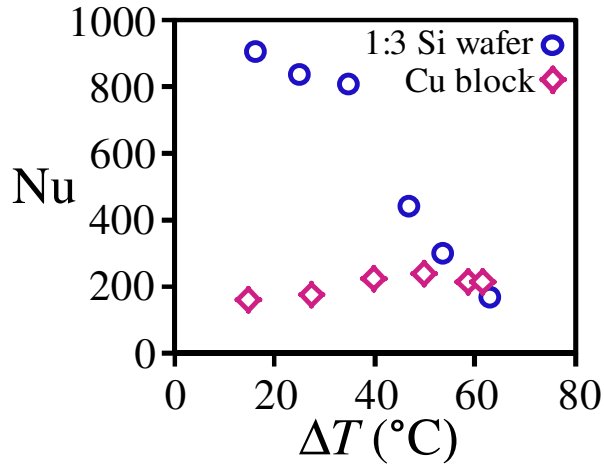

**Figure 7**. Heat transfer coefficients, expressed as Nusselt number (Nu), as a function of the degree of sub-cooling ($\Delta T$) on a pure copper surface where water undergoes a steamwise condensation, and through the wettability patterned silicon wafer pressed against the copper block, where water is continuously removed from the hydrophobic patches to the hydrophilic channels and then drained by gravity through the sides of the wafer.

While the main subject of this paper is to report that a diffusive flux of the condensed drops toward an absorbing boundary lead to a directed motion, here we provide some preliminary results to show how such a method of drop removal can be used in a thermal management technology.

Figure 7 compares the heat flux [represented as a Nusselt number $Nu = J_q D/(k_w \Delta T)$] as a function of the degree of sub-cooling $\Delta T$ under the conditions of a film-wise condensation on an uncoated copper and mixed dropwise and filmwise condensation on a striped silicon wafer pressed onto this copper substrate. It is evident that the heat transfer coefficient through the striped surface is considerably higher than the uncoated copper when $\Delta T$ is about 18°, in spite of the fact that there is an additional Kapitza resistance (due to differences in the free electron and phonon mediated heat transfers at the copper/silicon boundary) but the distinction disappears at much higher values of $\Delta T$, which is, of course, a non-practical regime for heat transfer applications It is interesting to note that the heat transfer coefficient (Nu~ 900) as observed here with a striped surface is very close to that (Nu~ 1000) observed previously, where condensed droplets were removed from a horizontal surface by a wettability gradient [4]. Since the operating range of $\Delta T$ of most heat exchangers is in the lower range, we can expect that a significant increase of the heat transfer coefficient can be achieved with a striped surface, on which the water drops condensed on the hydrophobic surface are removed by biased coalescence of the drops near the hydrophilic boundary.

**4. Conclusions.**

We demonstrated that droplets condensed on the hydrophobic region of a surface having alternate stripes of hydrophobic/hydrophilic zones move with substantial speeds towards the hydrophilic region of the surface. We believe that this motion results mainly from the diffusive flux of the coalesced drops from the hydrophobic to the hydrophilic sides of the surface although an in-situ produced thermal gradient due to differential heat transfer coefficients of the hydrophilic and hydrophobic stripes could provide additional driving force for such a motion. The motion can, however, occur even without a lateral thermal gradient. This method does not require the design of a pre-existing gradient, which is thus very useful in the passive removal of droplets from the steam side of a heat exchanger. The similarity between the phenomena described here and the diffusive elimination of a species between two absorbing boundaries is striking, except that the noise is athermal (but self-generated due to random coalescence) and the diffusion is non-linear in that it depends on the noise strength super-linearly. With these guidelines, we expect to develop a detailed statistical mechanical model of the phenomena reported here in near future. Further experiments will also be carried out to optimize the heat transfer rates by studying different ratios of the hydrophilic to hydrophobic area and using other types of non-trivial patterns.

**Appendix**

**A1**. **Demonstration of the noise induced coalescence and motion of drops on a surface having a radial gradient of wettability**.

A radial gradient was prepared by a diffusion controlled silanization process as described in reference 43. The diameter of the active zone of wettability is about 5 mm. The center is hydrophobic where the contact angle of water is about $110^o$, but the edges are wettable by water. When several drops are water are placed on such a surface, they do not move even though there is a wettability gradient. This kind of pinning of drop occurs because of wetting hysteresis. However, when the surface is vibrated vertically with a Gaussian random noise (at noise strength 0.1 $m^2/s^3$), drops begin to coalesce (as shown by circles) and move away from the center to the edge of the gradient. (See Movie 3 in Supporting Information)

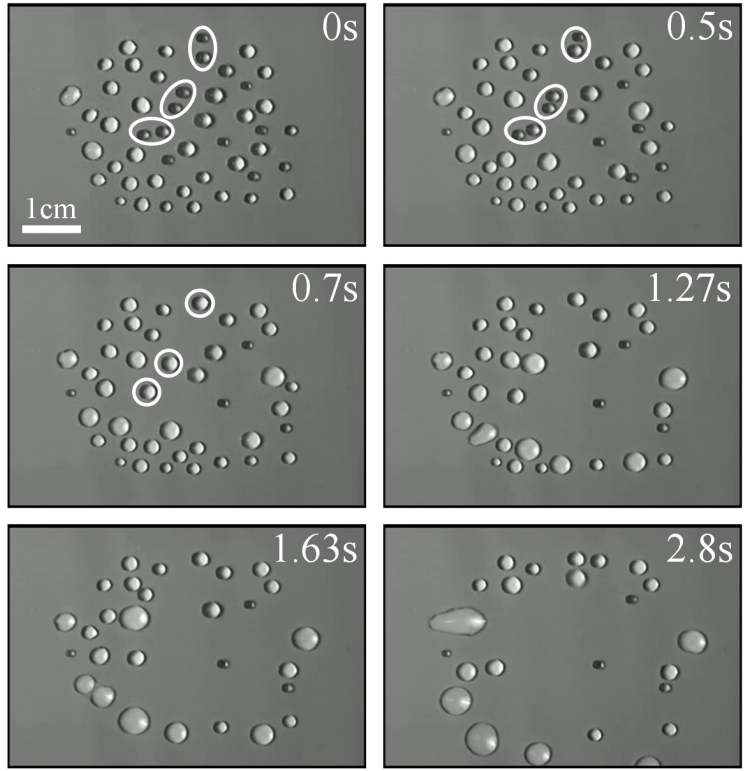

**Figure A1**. Demonstration of the noise induced coalescence (as shown by circles) and the motion of drops on a surface having a radial gradient of wettability. Noise strength is 0.1 m$^2$/s$^3$.

## A2. Similarity Between Noise and Coalescence Induced Motion of Drops on a Gradient Surface.

The drop motion experiments were carried out either using a linear surface tension gradient designed on the surface chemically or with the surface tension gradient induced on the drop by designing a thermal gradient on the solid support. The chemical gradient was prepared with a diffusion controlled silanization of a silicon wafer using decyltrichlorosilane. The wettability gradient thus prepared led to the value of $(d\cos\theta/dx)$ as 88 m$^{-1}$. The drop motion experiments [36] were carried out with a 10 $\mu l$ water drop using at a Gaussian noise of various strengths. The noise strength is defined as: $K=\langle\gamma^2(t)\rangle\tau_c$, where $\gamma(t)$ is the value (m/s$^2$) of the noise pulse and $\tau_c$ is its duration (40 $\mu$s). The drift velocity of the drop increases sub-linearly with the noise strength and saturates, which is consistent with the prediction of equation (2).

In a typical heat exchange, where dropwise condensation occurs, the heat flux is proportional to the rate which the drops are removed from the surface, because the mass flow rate of condensate times the enthalpy of condensation is the heat released from the steam that flows through the copper block. Various control experiments performed by us show that these two numbers are indeed comparable. Since the degree of sub-cooling is the control parameter that determines the rate of nucleation, the number density of nuclei and the rate of coalescence it can be viewed as playing the same role as the noise strength in the above experiment. The sub-linear growth [4] of the heat flux with respect to the $\Delta T$, has the same general trend as the noise induced drift of a single drop on a gradient surface (Figure A2).

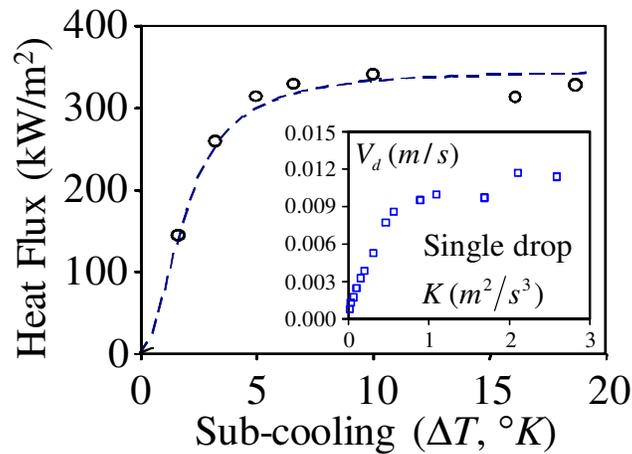

**Figure A2**. This figure shows the similarity between noise induced motion of a single drop (inset) on a surface having a wettability gradient [43] and the increase of the heat flux on a surface with the increase of $\Delta T$ – the degree of sub-cooling. Here, $V_d$ (m/s) is the drift velocity and $K$ (m$^2$/s$^3$) is the noise strength.

**A3. Role of Noise in Thermal Gradient Driven Motion of Liquid Drops**

Noise is also useful in inducing drop motion on a surface with athermal gradient, when it is thwarted by hysteresis. This was tested by designing a thermal gradient of 0.45 $^oC/mm$ on a uniform hydrophobic silicon wafer that is prepared by exposing it with the vapor of decyltrichlorosilane. The advancing and the receding contact angles of water on such a surface were 114$^o$ and 102$^o$ respectively. The drop motion experiments were carried out with a 4 $\mu l$ water drop using a Gaussian noise strength of 0.8 $m^2/s^3$. Here too, the drop would not move on the thermal gradient surface on its own. However, when the noise was introduced, the drop did move rather slowly (~ 1 mm/s) on the substrate. As the silicon wafer was inclined upward, the gravity opposed the motion of the drop. At a critical inclination, the drop stopped moving that yielded the estimate of $\bar{\gamma}$ as 0.11 $m/s^2$. The fact that noise overcomes hysteresis again is behind the reason why the condensing drops move on a surface discussed in this study. Here, random coalescence is the source of noise in the system.

**Acknowledgements**

This was supported by the Franklin J. Howes Jr. Distinguished Professorship to M. K. Chaudhury at the P. C. Rossin College of Engineering at Lehigh University.

**Supporting Information Available:** Three movies. Click on the links to view the movies.

Movie 1: Coalescence of drops near a hydrophilic boundary in heat transfer.

https://www.youtube.com/watch?v=ikbbUw096OE&list=UU5IDLAaiD_EQRe20_j5X7pA

Movie 2: Coalescence of drops near a hydrophilic boundary: drops generated on surface by mist of water.

https://www.youtube.com/watch?v=uNjfy_Z7aEU&list=UU5IDLAaiD_EQRe20_j5X7pA&index=2

Movie 3: Motion of drops on a surface having a radial gradient of wettability

https://www.youtube.com/watch?v=jWZz1MXC49M&list=UU5IDLAaiD_EQRe20_j5X7pA&index=1